\documentclass[11pt,a4paper]{scrartcl}

\usepackage[T1]{fontenc}
\usepackage[utf8]{inputenc}
\usepackage{textcomp}
\usepackage{amsmath, amssymb, amstext}
\usepackage{tikz,graphicx}
\usepackage{hologo}
\usepackage{pgfplots}
\usepackage{float}
\usepackage{calc}
\usepackage{subcaption}
\usepackage{titling}
\usepackage[export]{adjustbox}
\usepackage[a4paper]{geometry}
\usepackage{blindtext}
\usepackage{fancyhdr,graphicx,lastpage}
\usepackage{tabu}
\usepackage{caption}
\usepackage{threeparttable}
\usepackage{booktabs}
\usepackage{siunitx} %
\PassOptionsToPackage{hyphens}{url} %
\usepackage{hyperref} %
\usepackage{fancybox} %

\usepackage[
backend=biber,
style=numeric,
citestyle=verbose,
sorting=none
]{biblatex}

\usepackage{array}
\newcolumntype{L}[1]{>{\raggedright\let\newline\\\arraybackslash\hspace{0pt}}m{#1}}
\newcolumntype{C}[1]{>{\centering\let\newline\\\arraybackslash\hspace{0pt}}m{#1}}
\newcolumntype{R}[1]{>{\raggedleft\let\newline\\\arraybackslash\hspace{0pt}}m{#1}}

\usepackage[parfill]{parskip} %

\newcounter{romanpage}
\bibliography{Paper}

\fancypagestyle{plain}{
	\fancyhf{}
	
	\fancyhead[L]{\leftmark}
	\fancyhead[R]{David Knothe, Frederick Pietschmann}
	
	\fancyfoot[C]{\thepage}
	
}

\begin{document}
\pagestyle{plain}
\pagenumbering{Roman}

\newgeometry{top=2cm, bottom=1cm} %

\thispagestyle{empty} %

\begin{center}
	\vspace*{30mm}
	\vspace{10mm}
	{\Huge Large-Scale-Exploit of GitHub Repository Metadata and Preventive Measures\par}
	\vspace{25mm}
	{\LARGE David Knothe, Frederick Pietschmann\par}
	{\LARGE Karlsruhe: August 19, 2019 \par}
	{\LARGE Version 1.2 \par}
	
	\vspace{30mm}
	\doublebox{arXiv:1908.05354 [cs.CR]}
	\vspace{11mm}
\end{center}

\pagebreak

\restoregeometry %

\section*{Abstract}
\fancyhead[L]{}

When working with Git, a popular version-control system, email addresses are part of the metadata for each individual commit. When those commits are pushed to remotes, those email addresses become visible not only to fellow developers, but also to malicious actors aiming to exploit them.

Remote Git repositories are often hosted by companies specialized in doing so. By far the most popular and relevant of them is GitHub. While GitHub is seriously committed to protecting user data, resulting in tools like a \textit{noreply}-email-address service, it also provides an API that, as a side-effect, enables black-hat-hackers to easily fetch repository data.

As a part of our research we created a tool that leverages this API to collect and analyze user data, further referred to as \texttt{The Monster}. The data processed by it not only gives access to millions of email addresses, but is powerful and dense enough to create targeted phishing attacks posing a great threat to all GitHub users and their private, potentially sensitive data. Even worse, existing countermeasures fail to effectively protect against such exploits.

As a consequence and main conclusion of this paper, we suggest multiple preventive measures that should be implemented as soon as possible. We also consider it the duty of both companies like GitHub and well informed software engineers to inform fellow developers about the risk of exposing private email addresses in Git commits published publicly.
	
\pagebreak

\fancyhead[L]{}
\tableofcontents

\clearpage
\setcounter{romanpage}{\arabic{page}}

\pagebreak

\pagenumbering{arabic}
\fancyhead[L]{\leftmark}

\section{Introduction}
\label{introduction}

\subsection{Git}
\label{introduction-git}

Git is a distributed version-control system with a great focus on integrity. Primarily designed to track changes in source code,  it's still capable of handling any kind of file. Git was created by Linus Torvalds in 2005, when he was looking for an appropriate system to manage and coordinate the development of the Linux kernel. Git is open sourced and distributed under the terms of the GNU General Public License v2.

Being a distributed system, every Git repository on a computer stores the entire history, only requiring network access when fetching the repository from a remote or when pushing to a remote. A Git repository consists out of an arbitrary number of branches that may be merged into each other. Branches themselves are made out of a stack of commits and a reference to an originating commit. Commits store changes to files, newly added files, moved files, removed files as well as metadata like commit message, author and author date.

To enable such a system, commits are required to be uniquely identifiable. This is realized using a 40-digit-long, hexadecimal character-based SHA-hash (e.g. \textit{b76acbed\allowbreak ae520185e53\allowbreak 14d4973b4ed\allowbreak 30a4a5640f}) that takes multiple parameters as an input:\footcite{git-sha-hash}

\begin{itemize}
	\item Hash representation of the file tree
	\item Hashes of parent commits
	\item Commit message
	\item Author (name \& email)
	\item Author date
	\item Committer (name \& email)
	\item Commit date
\end{itemize}

\begin{figure}[!h]
	\centering
	\frame{\includegraphics[width=1\textwidth]{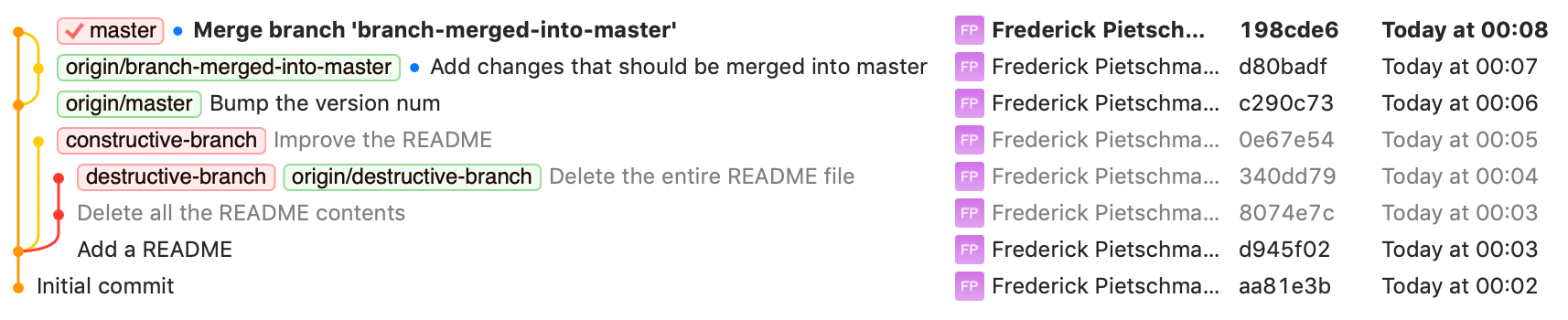}}
	\caption{Visual representation of a Git repository with multiple branches: one merged into the main branch (\textit{master}), two others not merged; one not pushed to \textit{origin} at all, one not up-to-date on \textit{origin}, one only available at \textit{origin}, one available both locally and at \textit{origin}. Also visible: Commit metadata including author name, commit hash and author date.}
	\label{fig-git}
\end{figure}

 As can be seen in figure \ref{fig-git}, Git clients often only display the first digits of such a hash. Still, all hashes have the same actual length of 40 hexadecimal digits.

\subsection{GitHub}

GitHub is a company independent from Git itself. As its main feature, GitHub provides hosting of Git repositories. While there exist some other popular services hosting Git projects, GitHub clearly leads the market. In addition to open source projects and personal projects, GitHub is also often used in an enterprise setting, by companies like Apple, Google or Microsoft, the latter of which acquired GitHub in 2018. Repositories hosted on GitHub may be openly visible or kept private and therefore only visible to those with the appropriate privileges.

GitHub not only offers basic Git hosting, but also many more features in conjunction with software development: Issue management, documentation support, wikis, statistics, CI integration, an API, allowing to fetch data hosted on GitHub in a convenient format, and advanced user management allowing organization-like structures and fine-grained, per-repository management of user privileges. Being a service able to manage software with such a rich feature set, GitHub must of course ensure security at a very high level. This is why, in 2013, GitHub launched its Bug Bounty program\footcite{github-bug-bounty}, allowing security researchers to privately submit vulnerabilities and earn a monetary and honorary reward.

\subsection{Idea}

As explained in section \ref{introduction-git}, users authoring or committing a commit on a Git repository will have to provide an email address that is taken into account among other data to calculate a commit hash. When this commit is pushed to a remote that's publicly accessible, anyone can simply collect this email address by cloning the repository and looking it up. This process can be automated and will then not only be able to just accumulate a multitude of email addresses, but merge the results of multiple such analyzed repositories together to create a database of personal profiles containing name, used email addresses and repositories committed to. 

It's this potentially large database that may then be abused for malicious purposes like phishing attacks. In this paper, we will show how simple it is to create such a database and what effects are to be expected when abused by a malicious actor.

\subsection{Motivation}

The authors of this paper are active and enthusiastic users of the services GitHub provides and most of their software development work leverages GitHub's rich feature set. We're aware how important it is to keep GitHub a secure place. This is why we want effective and appropriate countermeasures to be put in place, mitigating the threat of phishing attacks and alike based on collection and analysis of GitHub repository metadata.

In addition to describing the attack (chapter \ref{attack}) and its impact (chapter \ref{impact}), we therefore propose preventive measures (chapter \ref{countermeasures}) as the key takeaway from this paper.

\subsection{State of The Art}

\subsubsection{Previous Exploits}
\label{previous-exploits}

Attacks to collect email addresses from public Git repositories, mostly hosted on GitHub, aren't a new phenomenon at all. Here are some existing approaches:

\begin{description}

\item [GitHub API]
The Github API itself can be used to retrieve all users and get event information for each of these users using the \textit{https://api.github.com/\allowbreak users/\allowbreak\{username\}/\allowbreak events/public} url. This event information will only contain an actual email address if the user has not or just recently activated GitHub's \textit{noreply}-email-address service. Also, there exist API rate limits (see section \ref{existing-countermeasures}) that only allow a maximum of as few as 30 users per minute. Therefore, this approach is not feasible at all.

\item[GitHub \texttt{.patch} Links]
A commit to a repository has a unique link on GitHub. By appending the suffix \texttt{.patch} to this unique url, GitHub will return the raw commit diff (see figure \ref{fig-github-patch-link}), similar to the output of the \texttt{git-format-patch} command. This diff will also show the email address used by the author. While this approach works fine to get an email address for one specific commit, it's not scalable at all and doesn't offer significant advantages over cloning the repository and looking the changes up via Terminal or via a Git client.

\begin{figure}[!h]
	\centering
	\frame{\includegraphics[width=0.8\textwidth]{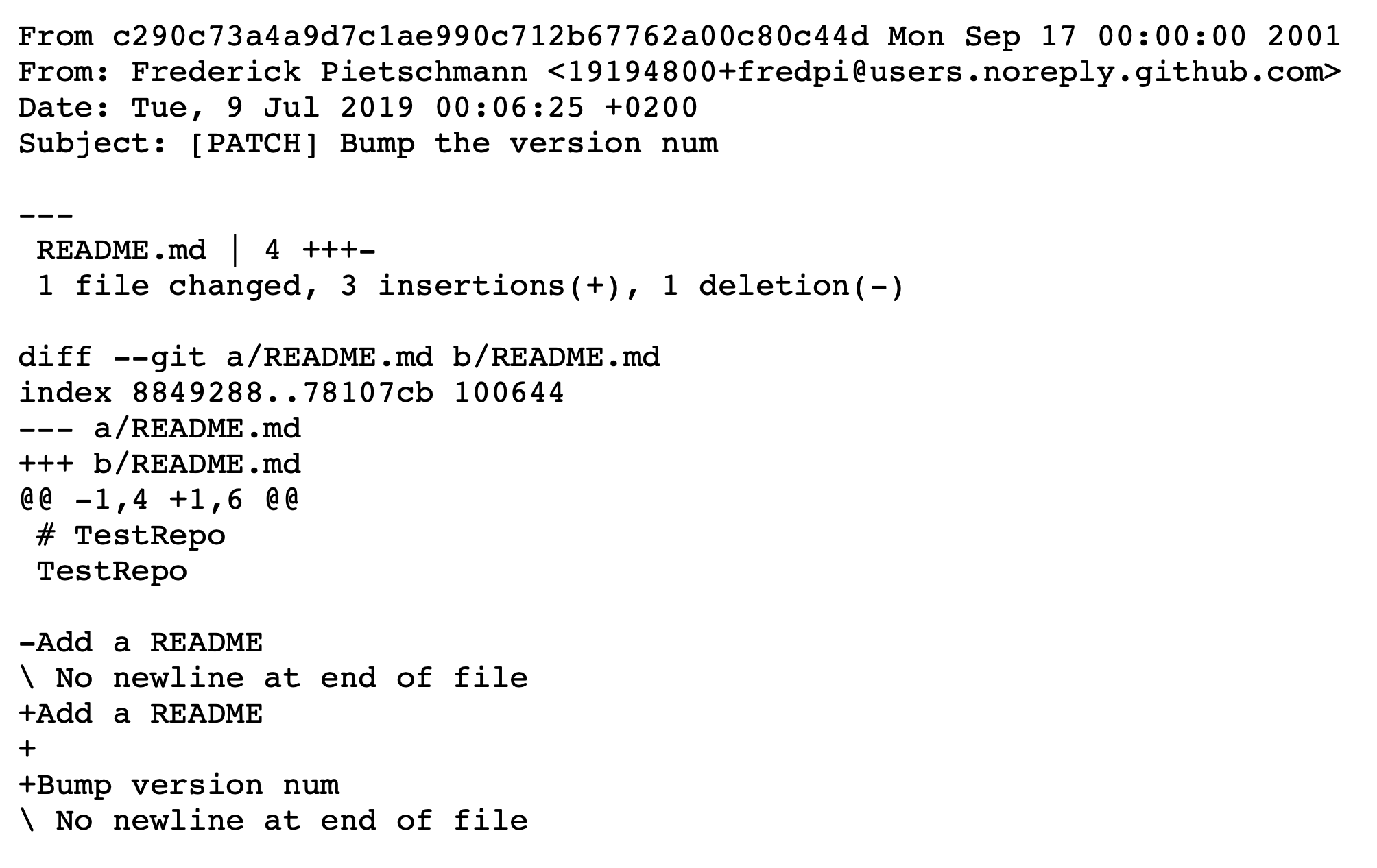}}
	\caption{Example output of a GitHub \texttt{.patch} link}
	\label{fig-github-patch-link}
\end{figure}

GitHub also recognizes the missing efficiency and scalability of this exploit and therefore explicitly declares this kind of approach an ineligible submission for the GitHub Bug Bounty Program:

\quote{There are a handful of reports that we consider ineligible, either because the feature is working as intended or we accept the low risk as a security/usability tradeoff.\footcite{github-bug-bounty-ineligible}}

\end{description}
\begin{description}

\item[All GitHub Commit Emails]
In 2016, a repository named \texttt{all-github-commit-emails}\footcite{all-github-commit-emails} appeared on GitHub, containing more than five million email addresses originating from GitHub commits. To collect all these email addresses, the white-hat hacker leveraged the GH Archive\footcite{github-archive}, a third party service caching GitHub's contents and making them accessible to the world via a convenient API.

The repository was quickly put down, but it's still documented how things evolved back in these days. As a consequence of this misuse of its API, GH Archive quickly agreed to introduce partial hashing for all email addresses to avoid further such attacks.\footcite{github-archive-email-fix}

Notably, this exploit didn't go further than just providing the raw email list, still, it was considered a threat and measures to make such a hack much harder were quickly put in place.

\end{description}

\subsubsection{Existing Countermeasures}
\label{existing-countermeasures}

As it's always the case with hackers / security researchers on the one side and developers on the other side, an exploit is often met with a countermeasure to mitigate or cancel the effect of the exploit. Existing preventive measures against the exploit of GitHub repository metadata shall be laid out in this section.

\begin{description}

\item [GitHub API Rate Limit]
To prevent misuse of its API, GitHub has limited the number of requests that can be made per timespan. Table \ref{table-rate-limit} shows the rate limits of the GitHub API as defined in its documentation.\footcite{github-api-rate-limiting}
	
\begin{table}[H]
\begin{center}
\begin{threeparttable}
\begin{tabular}{|c|c|c|}
	\midrule
	\textbf{Type} & \textbf{Regular API} & \textbf{Search API}\\
	\midrule
	Unauthenticated & 60 $\frac{Requests}{Hour}$ & 600 $\frac{Requests}{Hour}$\\
	\midrule
	Authenticated & 5000 $\frac{Requests}{Hour}$ & 1800 $\frac{Requests}{Hour}$\\
	\midrule
\end{tabular}
\caption{GitHub API rate limits}	
\label{table-rate-limit}
\end{threeparttable}
\end{center}
\end{table}
	
With these limits, the effect of approaches primarily relying on GitHub's API is severely mitigated, making them negligible. However, a user with enough knowledge, drive and patience will naturally be able to build a botnet circumventing these measures, still leaving the question whether such criminal action is worth the effort and risk of punishment.
		
\item [GitHub API Abuse Detection Mechanism]
		
In addition to the regular rate limits described above, there's also a dedicated abuse detection mechanism.\footcite{github-abuse-rate-limits} It's designed to detect bot- and spam-like behavior, while not interfering with appropriate API usage. How exactly this system is implemented is unclear, but it certainly won't be too restrictive given it's intended not to disturb regular users.

\item [Email Address Hashing]

As described in \ref{previous-exploits}, one email address exploit relied on data collected by GH Archive, an independent GitHub caching service. As a response, this caching service went to SHA1-hash the non-domain-component of all email addresses it caches.

While this measure shows the good will of this specific caching service, a malicious actor will still be able to build their own caching service or email address collection tool with the very same tools GH Archive used to acquire their data: The GitHub API and Git Cloning, followed by suitable processing and analysis. Actually, this paper, while not created with bad intent, exactly describes such an attack.
			
\item [GitHub Noreply-Email-Addresses]

In 2013, GitHub announced\footcite{github-private-emails-blog} a feature called \textit{Keep my email address private}. If a user opts to use this feature, all web-based Git operations, from then on, use a unique \textit{noreply}-email-address provided by GitHub instead of the primary email address provided by the user. A user could also locally set this email address as their commit email address, thereby preventing local Git operations from using and exposing their regular email address.

In 2017, GitHub introduced\footcite{github-more-private-emails-blog} an update to this feature, namely an option called \textit{Block command line pushes that expose my email}. It does exactly what its name says: If enabled, pushes that contain commits with a personal email address (more precisely: an email address associated with the user's GitHub account) will be blocked and therefore not be made available online. To fix this warning, one could e.g. rebase after configuring the \textit{noreply}-email-address, thereby dropping the private email address, then push again.

Those two features offer a pretty good protection against unwanted private email address exposure, but, as section \ref{monster-measurements} shows, only few people use them -- probably because these features are not enabled by default for new accounts. Also, previous commits utilizing a private email address cannot be updated to use the \textit{noreply}-email-address. Therefore, the \textit{noreply}-email-address approach is only fully effective when used starting with the very first commit of a user -- section \ref{monster-measurements} also shows the lacking effectivity of the \textit{noreply}-email-address approach when that's not the case.

\begin{figure}[!h]
	\centering
	\frame{\includegraphics[width=0.9\textwidth]{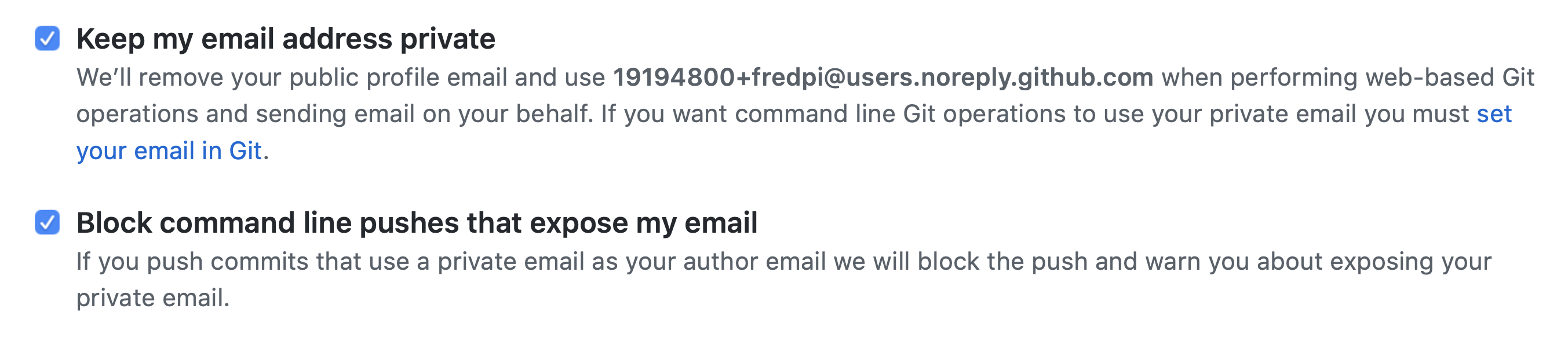}}
	\caption{The two GitHub settings allowing management of the \textit{noreply}-email-address}
	\label{fig-keep-email-private-block-pushes}
\end{figure}

\item [GitHub 2-Factor-Authentication]

GitHub optionally offers 2-Factor-Authentication.\footcite{github-two-factor-authentication} With this feature enabled, user must not only provide their password when logging in, but additionally confirm their identity via an independent device or service associated with their account. Therefore, even if a malicious actor may get access to user credentials via a targeted phishing attack, there's still the safeguard that they won't be able to log in without supplementary confirmation of the user.

In contrast to other countermeasures, this feature doesn't aim to defend the collection of data in the first place, but rather prevent account compromise even after a malicious actor already achieved to collect user credentials.
	
\end{description}

\pagebreak

\section{Attack}
\label{attack}

In this chapter we describe the attack: Using public GitHub repositories to collect data, analyze it to get detailed information about real persons and finally exploit it on a large scale. While this chapter describes the concept of the attack, chapter \ref{impact} will present measurements resulting from an actual implementation of the first part of this attack, thereby showing how serious this attack actually is and how simple and fast it is to gain a lot of useful data.

The following sections will address the respective methods and procedures of the three steps of the attack mentioned above: collecting, analyzing and exploiting data.

\subsection{Collecting Data}
\label{collecting-data}

In the first step of the attack -- \textit{Collecting Data} -- an arbitrarily large amount of public GitHub repositories will be cloned. This consists of two substeps -- fetching and cloning.

\subsubsection{Repository Fetching}

An attacker can retrieve information about all public repositories by using the official GitHub Search API. Simply performing a search for \texttt{stars:>=5} (\texttt{>=5} prevents an \textit{incomplete\_results} response) and sorting the result by number of stars will yield the top repositories -- up to 1,000 in a single search (spread over multiple pages). The resulting query url could look like this: \url{https://api.github.com/search/repositories?q=stars:>5\&sort=stars\&page=0}.

Retrieving more than 1,000 top repositories is comparably simple: once the first search for 1,000 repositories is exhausted, the next search string will be set to \texttt{stars:5..x}, where x is the lowest number of stars returned by the previous search. Using this process iteratively, an arbitrary number of top repositories can be fetched.

The data of each retrieved repository contains useful information: the repository name, the clone url, the main language (if present), the number of stars and the repository size. All this information can be stored and will be used later on (for cloning and analyzing).

\subsubsection{Repository Cloning}

After retrieving information about a repository, the attacker can clone the repository onto their local machine by performing  the \texttt{git clone} command.

To perform an efficient attack, one can determine which repositories should actually be cloned -- for instance, small repositories can be favored over larger repositories. A maximum size for cloning could also be set.

\subsection{Analyzing Data}
\label{analyzing-data}

\subsubsection{Single Repository}

After cloning a repository, the attacker can extract information about all the authors that have contributed to that repository. This is a feature of Git -- the \texttt{git shortlog -sne} command conveniently returns a list of all authors, alongside with their name, email address and respective number of commits.

\subsubsection{Person Matching}
\label{person-matching}

\begin{figure}[!h]
	\centering
	\includegraphics[width=0.65\textwidth]{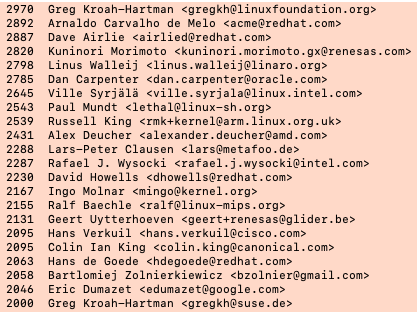}
	\caption{Example output of \texttt{git shortlog -sne} on \texttt{torvalds/linux}. At the very top and bottom, you see the same person using different email addresses, but the same name.}
	\label{fig-shortlog-example-output}
\end{figure}

Each line of the \texttt{shortlog}-output describes a person. Different lines may describe the \textbf{same} person (see figure \ref{fig-shortlog-example-output}). This is where \textit{person matching} -- the main point of the attack -- comes into play.

For example, a person could have made commits from different programs, computers or simply with different configurations. Then, they have commits with the same name, but a different email address, or vice versa.

This is detected by person matching. Two lines that are matching (same name or same email address) come from the same person. They can be merged into a single person data object.

\subsubsection{Multiple Repositories}

This gets particularly interesting when looking at multiple repositories. Many people contribute to more than one open-source project on GitHub. Therefore, when performing \texttt{git shortlog} on these repositories, the attacker will see one (or more) \texttt{shortlog}-lines of this person in each repository they have contributed to. These lines can \textbf{all} be merged into a single person object.

When doing this, the attacker will get the following information about a person: their name, all email addresses and all repositories they have committed to, tagged with the respective number of commits they have made.

This whole process is incredibly powerful when performed in an automated fashion:

\begin{itemize}
\item Clone a fixed (large) number of repositories and perform \texttt{git shortlog} on all of them.
\item Start with an empty list of person objects. Each object will contain information about a single real person.
\item Now, iterate over all \texttt{shortlog}-lines from all repositories. 
	\begin{itemize}
    \item If there exists a matching person object (having the same name or email address) in the list, merge the line into the existing person object.
    \item Otherwise, create a new person object just containing the information from this specific \texttt{shortlog}-line.
    \end{itemize}
\end{itemize}

At the end, there will be a full list of all involved persons. Besides the name and email addresses, each person object will contain a list of all repositories this person has contributed to.

Section \ref{monster-measurements} describes how fast it is to actually create such a full-fledged person database.

\subsubsection{GitHub Noreply-Email-Addresses}
\label{attack-noreply-email-addresses}

Later on, when explaining the exploits that can be performed with this database, all exploits will use the persons' email addresses in some way.

One may argue this whole attack is not problematic because there exist \textit{@users.noreply.\allowbreak github.com} email addresses. As explained in section \ref{existing-countermeasures}, these are email addresses issued by GitHub that are directly associated with a user's GitHub account. So, when committing, instead of using their real email address, a user can use the \textit{noreply}-email-address.

\begin{figure}[!h]
	\centering
	\includegraphics[width=0.9\textwidth]{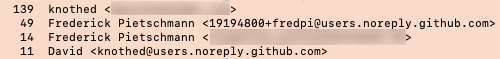}
	\caption{Example output of \texttt{git shortlog -sne} (taken from a real repository). Here, both persons used both their real email address and a \textit{noreply}-email-address. Both times, person matching detects that the lines belong to the same person; either because the real name is the same (Frederick Pietschmann), or even because the GitHub username matches the real name (knothed). \\ In this example, this was detected within a single repository -- these \texttt{shortlog}-lines could have just as well been scattered over multiple repositories, yielding the same result.}
	\label{fig-example-noreply}
\end{figure}

This does not stop the attacker from performing a successful attack, as \textit{noreply}-email-address addresses have two downsides:

\begin{enumerate}
	\item The user's GitHub name is literally inside the \textit{noreply}-email-address (e.g. \texttt{1024025+\allowbreak torvalds@users.noreply.github.com}). This allows the attacker to extract the GitHub username and use it later in an exploit.
	\item Sometimes, it happens that a person has some commits with a \textit{noreply}-email-address, but some other commits with their real email address (e.g. from a time where they did not use the \textit{noreply}-email-address yet, or simply via a commit from another computer or program where they accidentally didn't configure the \textit{noreply}-email-address). A single commit with the real email address is enough -- person merging will happily see that these \texttt{shortlog}-lines indeed have a different email address, but the same name (see figure \ref{fig-example-noreply}), and will therefore merge them together into one person object containing all data gathered up to this point.
\end{enumerate}

So, all commits that the user has made \textit{"anonymously"} -- without exposing their real email address -- can be exposed (linked to their real email address) just by a single, fatal commit that doesn't use the \textit{noreply}-email-address.

Section \ref{monster-measurements} describes how often this occurs.

\pagebreak

\subsection{Exploiting Data}
\label{exploiting-data}

Let's assume the attacker was able to retrieve a rich person database using the analysis step described above. In this section, we will show two possible ways they can exploit this data maliciously: spamming and phishing.

\subsubsection{Spamming}
\label{spamming}

This is a really simple attack. The attacker can use all email addresses from the database (either all or - probably better - just one per person) and use them for spamming, distributing a newsletter etc. They can leverage the IT-related background of the persons whose email addresses they were able to retrieve and adapt the content of the spam accordingly.

An attacker with enough dedication can even create targeted content for different persons up to the point where that content actually gets interesting for the targeted persons. That's possible because each person is associated with their contributed repositories, which themselves have a main programming language each, allowing content to adapt depending on the repositories or languages that a person regularly interacts with. Such a targeted approach could e.g. result in newsletters showing new trends or evolutions in the person's favorite programming language(s).

Performing this on a big scale (using all persons in the database) will yield at least some interested persons that will interact with the spam content -- may it be malicious (e.g. malware distribution) or just a newsletter advertising some product.

\subsubsection{Phishing}
\label{phishing}

Building on the concept of targeted email creation presented above, an attacker can trick users into trusting him, trying to impersonate GitHub itself to steal passwords from users.

Having gathered enough person data, the attacker is able to create authentic emails that look as they would originate from GitHub. Any link in such a mail would link to a phishing server replicating the GitHub login page. When the user actually enters their credentials there, the account is compromised and the attacker can  take control of the account (unless 2-factor-authentication is enabled).

Because the attacker has enough information about the person (real name, contributed repositories, favored programming languages), the person may actually be tricked into believing that the email is from GitHub itself. As this attack can simply be executed on all persons in the database, it would surely result in a significantly large number of victims.

\begin{paragraph}*{Email Creation}

Here, the attacker can be as creative as they want to be. Designing these phishing emails, it's certainly a good idea to leverage common ideas of social engineering, resulting in approaches like these:

\begin{itemize}
	\item Ask the user whether they really want to delete their repository (thereby inconspicuously requiring action from the user)
	\item Let the user know that they have been granted push access to a popular repository they previously committed to via pull requests (thereby abusing the curiosity of the user)
	\item Advertise repositories the user may like / display a notification for a repository the user has contributed to (thereby abusing trust originating from the authenticity of the email)
\end{itemize}

\end{paragraph}

\begin{figure}[H]
	\centering
	\frame{\includegraphics[width=0.7\textwidth]{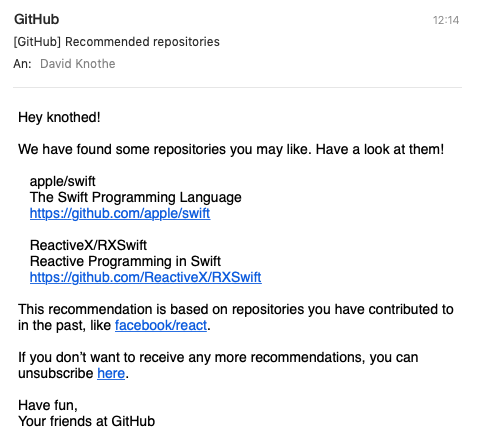}}
	\caption{A simple text-based phishing email with hyperlinks. The look is copied from subscription confirmation emails. The hyperlinks point to the phishing server, not to \url{https://www.github.com}. When the GitHub username isn't known, the person can be addressed with their real name or their given name (e.g. \textit{"Hey David"}).}
	\label{fig-phishing-mail-1}
\end{figure}

\begin{figure}[H]
	\centering
	\frame{\includegraphics[width=0.75\textwidth]{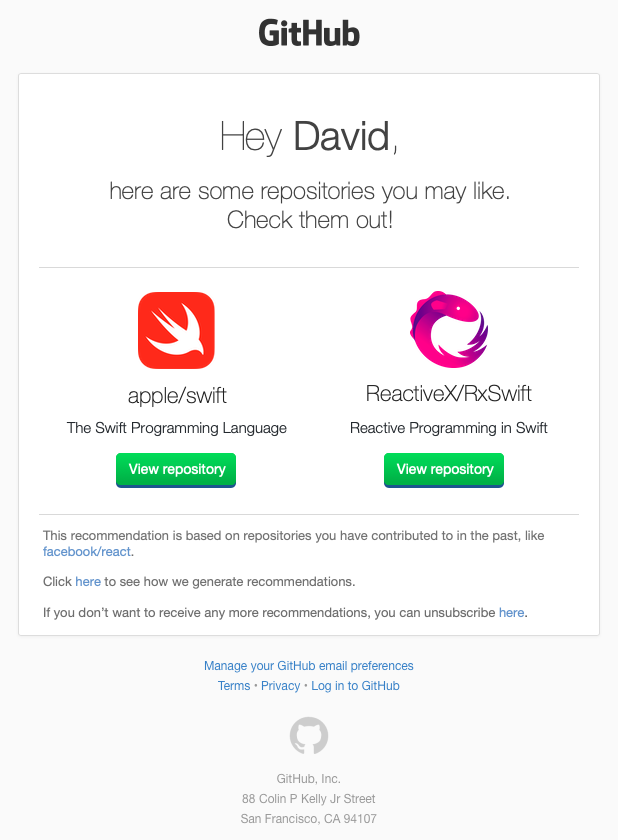}}
	\caption{An HTML-based phishing email. The look is copied from the \textit{"xyz has invited you to collaborate"} emails. Like above, all hyperlinks point to the phishing server.}
	\label{fig-phishing-mail-2}
\end{figure}

\begin{paragraph}*{Advertise Repositories} For the following section, we want to outline one of these approaches: Advertising repositories the user may like.

How does the attacker find such repositories? They create a graph describing the coherence between repositories. The vertices are all repositories they have analyzed. The edges are both weighted and directed and describe the coherence between pairs of repositories -- the higher the value of the edge, the more persons who have contributed to one repository have also contributed to the other repository.

At the beginning, there are no edges. Then, the attacker looks at each person who has contributed to more than one repository. They add a value (e.g. relative to the ratio of commit numbers) to all edges between pairs of repositories the user has contributed to. Doing this for all persons will yield a full graph where edges with higher values mean that persons who have contributed to one repository may also be interested in the other repository.

Now, when targeting a person, the attacker looks at their contributed repositories and leverages the graph created in the last step to conclude which repositories they may be most interested in. This is an intelligent use of social engineering -- building trust by using knowledge users only believe GitHub to have.

This whole process can be extended further by also taking the repositories' main programming languages into consideration -- a person is more likely to be interested in repositories that use the same programming language as repositories they have already contributed to.

After collecting this data, the last step is to actually create an email that looks like it comes from GitHub. Figures \ref{fig-phishing-mail-1} and \ref{fig-phishing-mail-2} show two example designs -- one text-based and one with an HTML-based look, both similar to emails that GitHub is actually sending.

\end{paragraph}

\pagebreak

\section{Impact} \label{impact}

Not only did we design the concept of the attack (chapter \ref{attack}), but also implemented it (apart from the actual exploitation component), thereby providing a proof-of-work. This actual implementation is referred to as \texttt{The Monster}. Using \texttt{The Monster}, it is incredibly simple to gather a large amount of data in little time. It allows us to make accurate measurements regarding speed and effectivity and to estimate these values for even larger datasets.

\texttt{The Monster} takes a number as its input: the number of top repositories (by stars) it should fetch. It then sorts the repositories by size and discards the largest 60\%. The remaining repositories will be cloned and analyzed in order to create a database with valuable user information.

This chapter presents measurements from multiple runs of \texttt{The Monster} and estimates the bad impact of \texttt{The Monster} when used by a malicious attacker.

\subsection{Theoretical Limits}

As \texttt{The Monster} is an implementation of the first steps of the attack described in chapter \ref{attack}, it both collects and analyzes data.

This section gives speed limits for these steps.

\begin{description}

\item [Repository Fetching]

This step is bounded by the GitHub Search API limit. The Search API allows up to 30 requests per minute (when authenticating the request with a real GitHub user). Each request returns a page with 100 repositories (by setting \texttt{per\_page=100}). This results in a limit of 3,000 repository fetches per minute.

Actual measurements from \texttt{The Monster} yield approximately 1,600 repositories per minute.

\item [Repository Cloning]

This step depends entirely on the attacker's network speed.
As described, to avoid cloning really large repositories, \texttt{The Monster} discards the largest 60\% of repositories before cloning.

This step is \textbf{not} bounded by the attacker's disk size: instead of cloning all repositories at once, \texttt{The Monster} uses an intelligent pipeline that clones and analyzes simultaneously in a producer/consumer fashion. For example, this guarantees that a maximum of 20 cloned repositories exists at the same time. After a repository has been cloned and analyzed, it will be deleted.

\item [Analyzing Data]

The \texttt{git shortlog} command is really fast -- for example, on \texttt{torvalds/\allowbreak linux} (a 3.5 GB large repository with > 800,000 commits) it takes around 30 seconds; on the majority of the repositories we tested, it takes well below one second.

The person matching process (described in section \ref{person-matching}) can be implemented intelligently using hash-based data structures to store which email addresses and names belong to which person object. Then, each merge of a new shortlog-line into the existing (possibly really large) person database runs in constant time (amortized). Therefore, the whole merge process takes a negligible time, similar to the \texttt{git shortlog} command.

\end{description}

\subsection{Monster Measurements}
\label{monster-measurements}

Of course, the main point of this attack is its large scale -- tons of repositories must be analyzed to create a large database. The following sections will show actual measurements from \texttt{The Monster} that originate from runs with different input values (\texttt{numRepos}).

\subsubsection{Timing}

As outlined in chapter \ref{attack}, \texttt{The Monster} does three things:
\begin{itemize}
	\item Fetching \texttt{numRepos} repositories
	\item Cloning the smallest 40\%\footnotemark \ of these repositories
	\item Analyzing (\texttt{shortlog} \& merge) these repositories.
\end{itemize}

\footnotetext{Why 40\%? This is just a value that seems to be a good fit. An attacker could try to optimize this parameter to further improve the efficiency of \texttt{The Monster}.\footnotemark}
\footnotetext{Actually, we first tried to sort the repositories by $\frac{\text{contributors}}{\text{size}}$ instead of the current approach ($\frac{\sqrt{\text{stars}}}{\text{size}}$), so the repositories with the best \textit{contributors per kilobyte ratio} were preferred. Sadly, both endpoints where the contributor count can be retrieved (1.: GitHub Repositories API, 2.: \texttt{/contributors\_size} route on the repository url) have good abuse detection mechanisms in place preventing \texttt{The Monster} from fetching this value on a large scale.}

\textit{Step 1}, fetching, is done first. When finished, \textit{step 2}, cloning and analyzing, begins. Cloning and analyzing is done simultaneously: each cloned repository is analyzed immediately.

The following table shows the measured data (all values in seconds). These measurements were done with a network speed of approximately 250 MBit/s.

\begin{table}[H]
  \begin{center}
  \begin{threeparttable}
  \caption{Timing data from \texttt{The Monster}}	
    \begin{tabular}{|c|c|c|cccc|}
    \toprule
      \texttt{numRepos} & \textbf{Total Time} & \textbf{Step 1} & \textbf{Step 2} & clone & \texttt{shortlog} & merge\\
      \midrule
      100 & 25.6 & 3.0 & 22.6 & 21.5 & 2.3 & 0.3\\
      300 & 46.1 & 10.7 & 35.4 & 34.0 & 3.9 & 0.7\\
      1,000 & 130.3 & 29.9 & 100.4 & 96.3 & 13.2 & 2.2\\
      3,000 & 391.6 & 122.5 & 269.1 & 258.2 & 30.4 & 6.5\\
      10,000 & 1,138 & 360.6 & 777.6 & 746.7 & 84.8 & 23.8\\
      30,000 & 3,211 & 1,226 & 1,984 & 1,922 & 254.9 & 68.7\\
      55,000 & 5,334 & 1,916 & 3,418 & 3,352 & 345.4 & 116.4\\
      \bottomrule
    \end{tabular}
    \begin{tablenotes}
      \small
      \item clone, \texttt{shortlog} and merge added up take longer than step 2 -- this is because cloning, \texttt{shortlog} and merging are performed simultaneously.
    \end{tablenotes}
  \label{table-speed}
  \end{threeparttable}
  \end{center}
\end{table}

As you can see, \texttt{totalTime} is approximately linear in \texttt{numRepos} with a factor of $0.1$. Because cloning takes the most time, this number strongly depends on the network speed and is only minimally dependent on the machine speed (which is only relevant for \texttt{shortlog} and merge).

\subsubsection{Database Size}

Here, we show statistics of the generated database. This includes
\begin{itemize}
	\item The number of failed repositories. This means that an error occurred either on cloning or analyzing
	\item The total number of persons in the database. This is the most relevant value as it describes the total size of the generated database
	\item Persons with more than one repository. This is relevant for the repository coherence graph (see section \ref{phishing})
	\item Persons with 5 or more commits in total
	\item The total number of \texttt{shortlog}-lines from all repositories.
\end{itemize}

\begin{table}[H]
  \begin{center}
  \begin{threeparttable}
  \caption{Database statistics from \texttt{The Monster}}	
    \begin{tabular}{|c|c|c|c|c|c|}
    \midrule
      \texttt{numRepos} & \textbf{Failed} & \textbf{Total Persons} & \textbf{$\geq 2$ repos} & \textbf{$\geq 5$ commits} & \textbf{\texttt{shortlog}-lines}\\
      \midrule
      100 & 0.0\% & 15,880 & 6.1\% & 8.5\% & 18,130\\
      300 & 0.8\% & 24,435 & 10.2\% & 10.1\% & 29,891\\
      1,000 & 1.0\% & 44,041 & 15.0\% & 12.6\% & 59,305\\
      3,000 & 1.3\% & 71,917 & 20.0\% & 15.6\% & 107,853\\
      10,000 & 1.2\% & 116,883 & 25.3\% & 19.8\% & 201,814\\
      30,000 & 0.7\% & 169,961 & 29.4\% & 24.2\% & 347,321\\
      55,000 & 0.9\% & 210,164 & 31.6\% & 27.2\% & 463,130\\
	\midrule
    \end{tabular}
  \end{threeparttable}
  \end{center}
\end{table}

Astonishingly, only fetching the top 1,000 repositories (and cloning and analyzing 40\% of them) already yields a database containing 44,000 persons. As can be seen in table \ref{table-speed}, this only takes about two minutes!

\begin{figure}[H]
	\centering
	\includegraphics[width=0.8\textwidth]{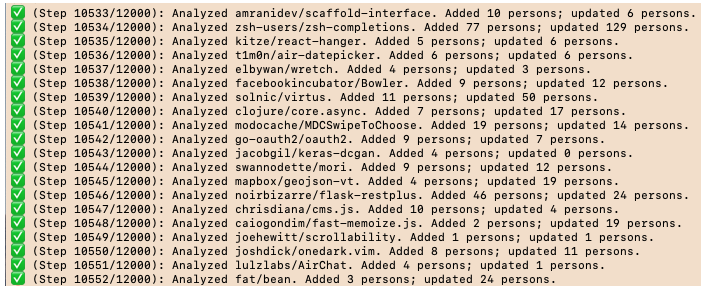}
	\caption{Output from \texttt{The Monster} with \texttt{numRepos=30000}. After having analyzed 10,500 repositories, the database is already fairly large causing many \texttt{shortlog}-lines to be matched and merged with existing persons. For instance, on \texttt{solnic/virtus} (step $10,539$), only 11 new persons were added, while 50 existing persons were updated.}
	\label{fig-monster-example-output}
\end{figure}

As expected, when analyzing more and more repositories, more and more persons are found to be contributing to at least 2 repositories (see figure \ref{fig-monster-example-output}). This explains why the growth of \texttt{totalPersons} is decreasing: the more persons are in the database, the more \texttt{shortlog}-lines (from a new repository) belong to an already existing person and therefore don't represent a new person.

The total number of \texttt{shortlog}-lines grows better, but, again, not linear. This is probably because we only look at the smallest 40\% of repositories. Changing this parameter may change this behavior -- we did not investigate this further.

\subsubsection{GitHub Noreply-Email-Addresses}

As described in section \ref{attack-noreply-email-addresses}, persons can use \textit{noreply}-email-addresses if they do not want their private email address to be publicly visible. Nevertheless, the analysis process may be able to match a \textit{noreply}-email-address to a private email address, for example when the person has made commits with different email addresses but using the same name or using their GitHub name.

\texttt{The Monster} does detect how often this happens. When a person uses a \textit{noreply}-email-address, \texttt{The Monster} stores their GitHub username (which is contained in the address) in the person data object. Therefore, we know that:

\begin{itemize}
	\item Every person in the database that has an associated GitHub username has definitely used a \textit{noreply}-email-address (there is no other way for \texttt{The Monster} to get the GitHub username of a person)
	\item A person with a GitHub username that \textbf{also} has an associated real (not \textit{noreply}) email address was \textit{"compromised"} in the sense that their private email address was exposed despite having the \textit{noreply}-email-address enabled.
\end{itemize}

The following table shows these two values: First, how many persons have \textit{noreply}-email-addresses enabled, and second, how many of those were compromised, meaning where an additional email address was found.

\begin{table}[H]
  \begin{center}
  \begin{threeparttable}
  \caption{Statistics about \textit{noreply}-email-addresses}	
    \begin{tabular}{|c|c|c|}
    \midrule
      \texttt{numRepos} & \textbf{\textit{noreply} enabled} & \textbf{still compromised}\\
      \midrule
      100 & 13.9\% & 8.3\%\\
      300 & 13.5\% & 13.1\%\\
      1,000 & 12.7\% & 16.7\%\\
      3,000 & 12.2\% & 20.9\%\\
      10,000 & 11.7\% & 26.3\%\\
      30,000 & 11.7\% & 31.6\%\\
      55,000 & 11.9\% & 34.5\%\\
      \midrule
    \end{tabular}
  \end{threeparttable}
  \end{center}
\end{table}

\begin{figure}[H]
	\centering
	\frame{\includegraphics[width=0.9\textwidth]{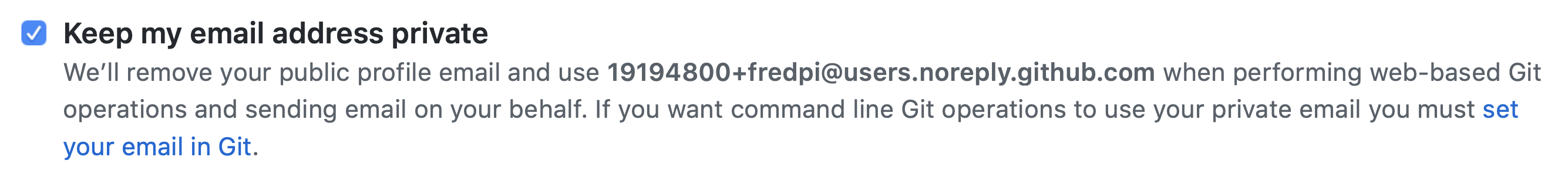}}
	\caption{Not as private as one may think}
	\label{fig-keep-email-private}
\end{figure}

As you can see, the more repositories we analyzed, the more \textit{noreply}-email-addresses (figure \ref{fig-keep-email-private}) could be compromised.

\subsection{Monster Estimates}
\label{monster-estimates}

As seen in the last section, an attacker can build a database containing over 200,000 persons in just 1.5 hours.

Of course, we do not know how far this method can be extended meaningfully. Currently, there exist 1.5 million public repositories with at least 5 stars, and 7 million public repositories with at least 1 star (as indicated by Search API requests).

For values up to 7 million, we can try to estimate the impact using our measurements from section \ref{monster-measurements}. The following graphs are all based on these measurements and equipped with sensible regression curves.

\subsubsection{Timing}

As the graphs below show, both the time for step 1 and the time for step 2 (therefore also the total time) behave approximately linear in the \texttt{numRepos} parameter.

\begin{figure}[H]  
\centering
\begin{subfigure}[b]{0.4\linewidth}
\begin{tikzpicture}
\begin{axis}[
   scale=0.8,
   xlabel={\texttt{numRepos} [$\times$ 1,000]},
    ylabel={Time [$\times$ 1,000 sec]},
    xmin=0, xmax=65,
    ymin=0, ymax=4,
    xtick={0,10,20,30,40,50,60},
    ytick={0,1,2,3,4},
    legend pos=north west,
    ymajorgrids=true,
    grid style=dashed,
]
 
\addplot[
    color=orange,
    mark=halfcircle*,
    mark size=3pt
    ]
    coordinates {
    (0.1,0.0003)(0.3,0.01)(1,0.03)(3,0.12)(10,0.36)(30,1.23)(55,1.91)
    };
    \legend{Step 1}
    
\addplot[
    color=green,
    mark=halfcircle*,
    mark size=3pt
    ]
    coordinates {
    (0.1,0.023)(0.3,0.035)(1,0.1)(3,0.27)(10,0.78)(30,1.99)(55,3.42)
    };
    \addlegendentry{Step 2}

\end{axis}
\end{tikzpicture}
\caption{Step 1 and step 2} \label{fig:M2}
\end{subfigure} \hskip 40pt
\begin{subfigure}[b]{0.4\linewidth}
\begin{tikzpicture}
\begin{axis}[
	scale=0.8,
   	xlabel={\texttt{numRepos} [$\times$ 1,000]},
    xmin=0, xmax=65,
    ymin=0, ymax=6.5,
    xtick={0,10,20,30,40,50,60},
    ytick={0,1,2,3,4,5,6},
    legend pos=north west,
    ymajorgrids=true,
    grid style=dashed,
]
 
\addplot[
	only marks,
    color=blue,
    mark=halfcircle*,
    mark size=3pt
    ]
    coordinates {
    (0.1,0.0026)(0.3,0.046)(1,0.13)(3,0.39)(10,1.14)(30,3.21)(55,5.33)
    };
    \legend{Total Time}
    
\addplot[
    domain=0:100, 
    samples=10, 
    color=red,
    dashed
]
 {0.1*x};

\end{axis}
\end{tikzpicture}
\caption{Total time with linear fit line} \label{fig:M2}  
\end{subfigure}
\end{figure}

Looking at \texttt{totalTime}, we have a relation of $\texttt{totalTime} \approx 0.1 \text{sec} \times \texttt{numRepos}$. Naturally, this factor depends on the network speed. Extending this further yields:

\begin{table}[H]
  \begin{center}
  \begin{threeparttable}
    \begin{tabular}{|c|c|}
    \midrule
      \texttt{numRepos} & \textbf{Total Time}\\
      \midrule
      100,000 & 2.8h\\
      1,000,000 & 27.8h\\
      7,000,000 & 194h\\
      \midrule
    \end{tabular}
  \end{threeparttable}
  \end{center}
\end{table}

An attacker with a higher network speed could generate even lower values.

\subsubsection{Database Size}

Here, the relation is not so obvious. One may think that, because the total time behaves linearly in respect to \texttt{numRepos}, the number of \texttt{shortlog}-lines should do so as well. But this isn't the case -- the curve rather looks like a square root curve. A good fit is $\texttt{shortlogLines} \approx \text{2,000} \times \texttt{numRepos}^{0.5}$.

Similarly, \texttt{totalPersons} can be approximated plausibly using a power curve:

$\texttt{totalPersons} \approx \text{3,310} \times \texttt{numRepos}^{0.382}$.

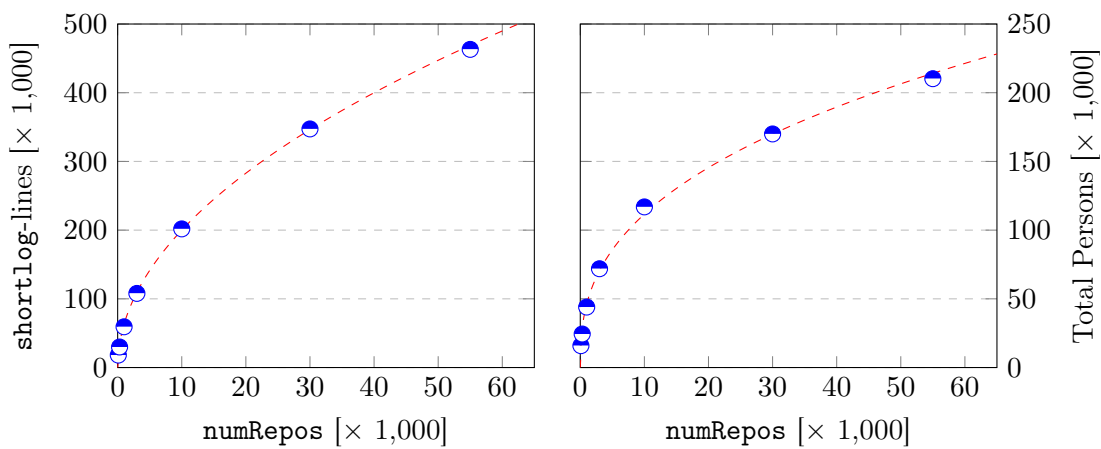
\begin{figure}[H]  
\centering
\begin{subfigure}[b]{0.4\linewidth}
\begin{tikzpicture}
\begin{axis}[
	scale=0.8,
   	xlabel={\texttt{numRepos} [$\times$ 1,000]},
   	ylabel={\texttt{shortlog}-lines [$\times$ 1,000]},
    xmin=0, xmax=65,
    ymin=0, ymax=500,
    xtick={0,10,20,30,40,50,60},
    ytick={0,100,200,300,400,500},
    legend pos=north west,
    ymajorgrids=true,
    grid style=dashed
]
 
\addplot[
	only marks,
    color=blue,
    mark=halfcircle*,
    mark size=3pt
    ]
    coordinates {
    (0.1,18.1)(0.3,29.9)(1,59.3)(3,107.9)(10,201.8)(30,347.3)(55,463.1)
    };
    
\addplot[
    domain=0:100, 
    samples=300, 
    color=red,
    dashed
]
{2.000*(1000*x)^0.5};

\end{axis}
\end{tikzpicture}
\end{subfigure} \hskip 40pt
\begin{subfigure}[b]{0.4\linewidth}
\begin{tikzpicture}
\begin{axis}[
	scale=0.8,
   	xlabel={\texttt{numRepos} [$\times$ 1,000]},
   	ylabel={Total Persons [$\times$ 1,000]},
    xmin=0, xmax=65,
    ymin=0, ymax=250,
    xtick={0,10,20,30,40,50,60},
    ytick={0,50,100,150,200,250},
    legend pos=north west,
    ymajorgrids=true,
    ylabel near ticks,
    yticklabel pos=right,
    grid style=dashed,
]
 
\addplot[
	only marks,
    color=blue,
    mark=halfcircle*,
    mark size=3pt
    ]
    coordinates {
    (0.1,15.9)(0.3,24.4)(1,44)(3,71.9)(10,116.9)(30,170)(55,210.2)
    };
    
\addplot[
    domain=0:100, 
    samples=300, 
    color=red,
    dashed
]
 {3.310*(1000*x)^0.382};   

\end{axis}
\end{tikzpicture}
\end{subfigure}
\caption{Total persons and total \texttt{shortlog}-lines, both with fit curves} \label{fig:M2}  
\end{figure}

Also, the \textit{persons-per-repo} ratio is decreasing quite fast. We don't exactly know why this happens, but it definitely has to do with the sorting. Instead of sorting the repositories by size before discarding 60\%, sorting by $\frac{\text{contributors}}{\text{size}}$ would give better results, because the most efficient repositories, in terms of people per time, would be picked.

Extending the \texttt{totalPersons} curve using $\texttt{totalPersons} \approx \text{3,310} \times \texttt{numRepos}^{0.382}$ yields:

\begin{table}[H]
  \begin{center}
  \begin{threeparttable}
    \begin{tabular}{|c|c|}
    \midrule
      \texttt{numRepos} & \textbf{Total Persons}\\
      \midrule
      100,000 & 269,000\\
      1,000,000 & 648,000\\
      7,000,000 & 1,364,000\\
      \midrule
    \end{tabular}
  \end{threeparttable}
  \end{center}
\end{table}

As stated in section \ref{previous-exploits}, a repository named \texttt{all-github-commit-emails} accumulated over 5 million committer email addresses. Because the authors of this repository just collected email addresses without performing the person matching process as described in section \ref{person-matching}, we should instead compare their result of 5 million email addresses against our \texttt{shortlogLines} measurements. When extending our regression curve ($\texttt{shortlogLines} \approx \text{2,000} \times \texttt{numRepos}^{0.5}$) to $\texttt{numRepos} = 7,000,000$, we get: $\texttt{shortlogLines} = 5,300,000$, matching the extent of the \texttt{all-github-commit-emails} exploit in terms of raw email address data.

We also believe that the expected value for \texttt{totalPersons} (1,364,000) is only a lower bound: when looking at much smaller and less known repositories (e.g. with less than 5 stars), there are much more \textit{"mavericks"} that just committed to their own repositories, not contributing to other open-source projects. This would mean that the ratio $\frac{\texttt{totalPersons}}{\texttt{shortlogLines}}$ would get a little bit bigger, making the estimate of only 1,364,000 just a lower bound, with higher values to actually expect.

\subsubsection{GitHub Noreply-Email-Addresses}

The percentage of persons that use \textit{noreply}-email-addresses stays constant at around 12\%.

For the percentage of compromised persons (persons using \textit{noreply}-email-addresses, where \texttt{The Monster} nonetheless found a private email address), we can look at the base-10-logarithm of \texttt{numRepos} to retrieve an approximate linear relationship. Looking at the graph below, we see: $\texttt{compromised-percentage} \approx 9.4 \times \text{log}_{10}\ \texttt{numRepos} - 11$.

\begin{figure}[H]  
\centering
\begin{tikzpicture}
\begin{axis}[
	scale=0.8,
   	xlabel={$\text{log}_{10}\ \texttt{numRepos}$},
   	ylabel={Compromised [\%]},
    xmin=1, xmax=5,
    ymin=0, ymax=40,
    xtick={1,1.5,2,2.5,3,3.5,4,4.5,5},
    ytick={0,10,20,30,40},
    legend pos=north west,
    ymajorgrids=true,
    grid style=dashed
]
 
\addplot[
	only marks,
    color=blue,
    mark=halfcircle*,
    mark size=3pt
    ]
    coordinates {
    (2,8.3)(2.5,13.1)(3,16.7)(3.5,20.9)(4,26.3)(4.5,31.6)(4.75,34.5)
    };
    <
\addplot[
    domain=0:5,
    samples=10,
    color=red,
    dashed
]
{9.4*x - 11};

\end{axis}
\end{tikzpicture}
\end{figure}

Extending this relation yields:

\begin{table}[H]
  \begin{center}
  \begin{threeparttable}
    \begin{tabular}{|c|c|}
    \midrule
      \texttt{numRepos} & \textbf{compromised}\\
      \midrule
      100,000 & 36\%\\
      1,000,000 & 45\%\\
      7,000,000 & 53\%\\
      \midrule
    \end{tabular}
  \end{threeparttable}
  \end{center}
\end{table}

In total, we see: an attacker can use a single day of computing power to generate a comprehensive database of 650,000 persons, or even use a full week of computing power to produce a twice-as-big database.

Out of the 12\% of persons that have enabled \textit{noreply}-email-addresses, the attacker knows their private email address in around 50\% (!) of the time.

\subsection{Phishing Estimates}

In sections \ref{monster-measurements} and \ref{monster-estimates} we presented actual measurements and precise estimates supporting our claim that the attack described in this paper is indeed dangerous to a certain extent. These values only described the effectivity of collecting and analyzing data. For a successful attack, the task of exploiting this data must also succeed to some degree.

Section \ref{exploiting-data} primarily focused on phishing as an effective way to exploit the data collected previously. Therefore, this section correspondingly focuses on providing estimates for the effectivity of a phishing attack as described in \ref{phishing}. As it's quite hard to determine precise numbers describing the success of a phishing attack without performing it, we only provide rather rough estimates based on existing research.

According to two sources \footcite{links-in-phishing-mails} \footcite{impact-usability-phishing}, between 8.8\% and 12\% of people click on a link in (non-targeted) phishing emails. Out of these, half of them provide their credentials to the phishing server.

In contrast, this source\footcite{phishing-attack-academic-context} investigated the effectivity of targeted phishing emails in an academic context. The researchers sent university-specific emails to college students. The results were astonishing: About 70\% of students clicked the link in the phishing email. There was a discrepancy between different majors: From all IT students, only about 35\% clicked the link. Because the target group of a phishing attack as described in section \ref{phishing} are mostly IT-related persons, we could assume a percentage not larger than these 35\%.

In total, we have a very vague estimate: between 8\% and 35\% of people click on links in phishing emails. The more tailored the phishing emails are, the higher the percentage, obviously. How good a phishing attack based on data collected by \texttt{The Monster} will perform is unclear, but its success ratio most likely lies somewhere between these bounds.

\pagebreak

\section{Countermeasures}
\label{countermeasures}

As chapter \ref{impact} clearly showed, the impact of \texttt{The Monster} and similar attacks must not be underestimated. It's not to be questioned that there have to be preventive measures effectively tackling this problem.

\subsection{Effectivity of Existing Countermeasures}
\label{effectivity-existing-countermeasures}

Section \ref{existing-countermeasures} presented countermeasures that are already set up. Those are:

\begin{itemize}
	\item GitHub API Rate Limit
	\item GitHub API Abuse Detection Mechanism
	\item Email Address Hashing
	\item GitHub Noreply-Email-Addresses
	\item GitHub 2-Factor-Authentication
\end{itemize}

All of these countermeasures aren't enough to prevent \texttt{The Monster} from succeeding:

\begin{itemize}
	\item The GitHub API rate limits and abuse detection mechanisms apparently don't prevent \texttt{The Monster} from fetching the top repositories via the GitHub API. As all further activity of \texttt{The Monster} doesn't leverage the GitHub API, the API rate limits and abuse detection mechanisms can't stop The Monster from then on.
	\item \texttt{The Monster} doesn't rely on third-party caching services like GH Archive. As such, it doesn't care whether email addresses stored in these caching services are hashed or not.
	\item As the numbers presented in sections \ref{monster-measurements} and \ref{monster-estimates} clearly show, only 12\% of the users collected by \texttt{The Monster} use GitHub's \textit{noreply}-email-address service. Out of them, only about 60\% are effectively protected via it, as for the other 40\%, a private email address could still be found somewhere in the history.
	\item We don't have official numbers, but expect the percentage of accounts that have GitHub's 2-Factor-Authentication feature enabled to be relatively low, as is the case with other services. For instance, only 10\% of Gmail accounts used Google's 2FA service in 2018.\footcite{two-factor-authentication-google} Even worse, the people already using 2FA are probably those that are security-aware and therefore won't even be tricked by a phishing attack in the first place. This is why, although 2-Factor-Authentication definitely acts as an effective protection for the ones who use it, it would probably not significantly mitigate the number of successfully compromised accounts if \texttt{The Monster} were to exploit the data it collected.
\end{itemize}

\subsection{Proposals}

As can be seen in section \ref{effectivity-existing-countermeasures}, the effectivity of the existing countermeasures is miserable. This is why there's a strong need for better, more successful approaches to defend against attacks similar to \texttt{The Monster}.

The responsibility to protect against malicious actors is not only with \textbf{Git} itself and services like \textbf{GitHub}. It's also a personal responsibility of each individual \textbf{User}. Generally speaking, it's very much needed to raise awareness in the developer community that this issue exists and how it may be misused. 

For these three domains (Git, GitHub and User) we want to propose solutions, most of which won't totally fix the problem once and forever, but at least mitigate it to some extent.

\subsubsection{Git}

While the original Git developers were the ones to make the decision to identify commit authors and committers not only by their name, but also by an email address they own or at least claim to own, they are not to blame for this decision. Back in the day, no one could have estimated how Git would evolve, especially that centralized platforms like GitHub play such a big role nowadays.

Having said that, it's nonetheless impossible to fundamentally change Git's design, even if it's considered flawed in that regard from today's perspective. The only solution would be to create an all-new Git version that entirely rethinks author identity management. As such a version would definitely break compatibility with existing Git versions that heavily rely on author name and email address for commit hashing (as pointed out in section \ref{introduction-git}), developing it is certainly off the table.

\begin{figure}[H]
	\centering
	\includegraphics[width=0.6\textwidth]{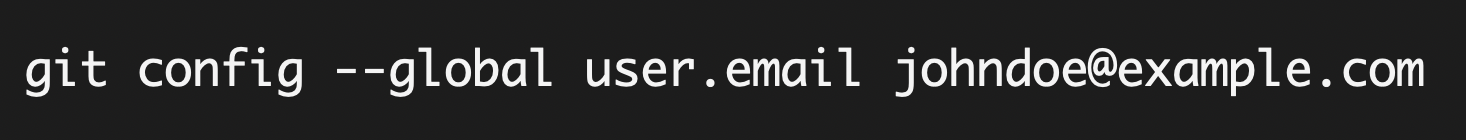}
	\caption{Performing such commands should cause a warning that the configured email address is recommended to be a dedicated commit email address}
	\label{fig-git-config-command}
\end{figure}

Therefore, the only thing that Git itself can do is raising awareness. For example, this can be done by pointing out that users shouldn't enter private email addresses when they are configuring their Git email address using commands like \texttt{git config} as can be seen in figure \ref{fig-git-config-command}.

\pagebreak

\subsubsection{GitHub}

\begin{figure}[H]
	\centering
	\includegraphics[width=0.9\textwidth]{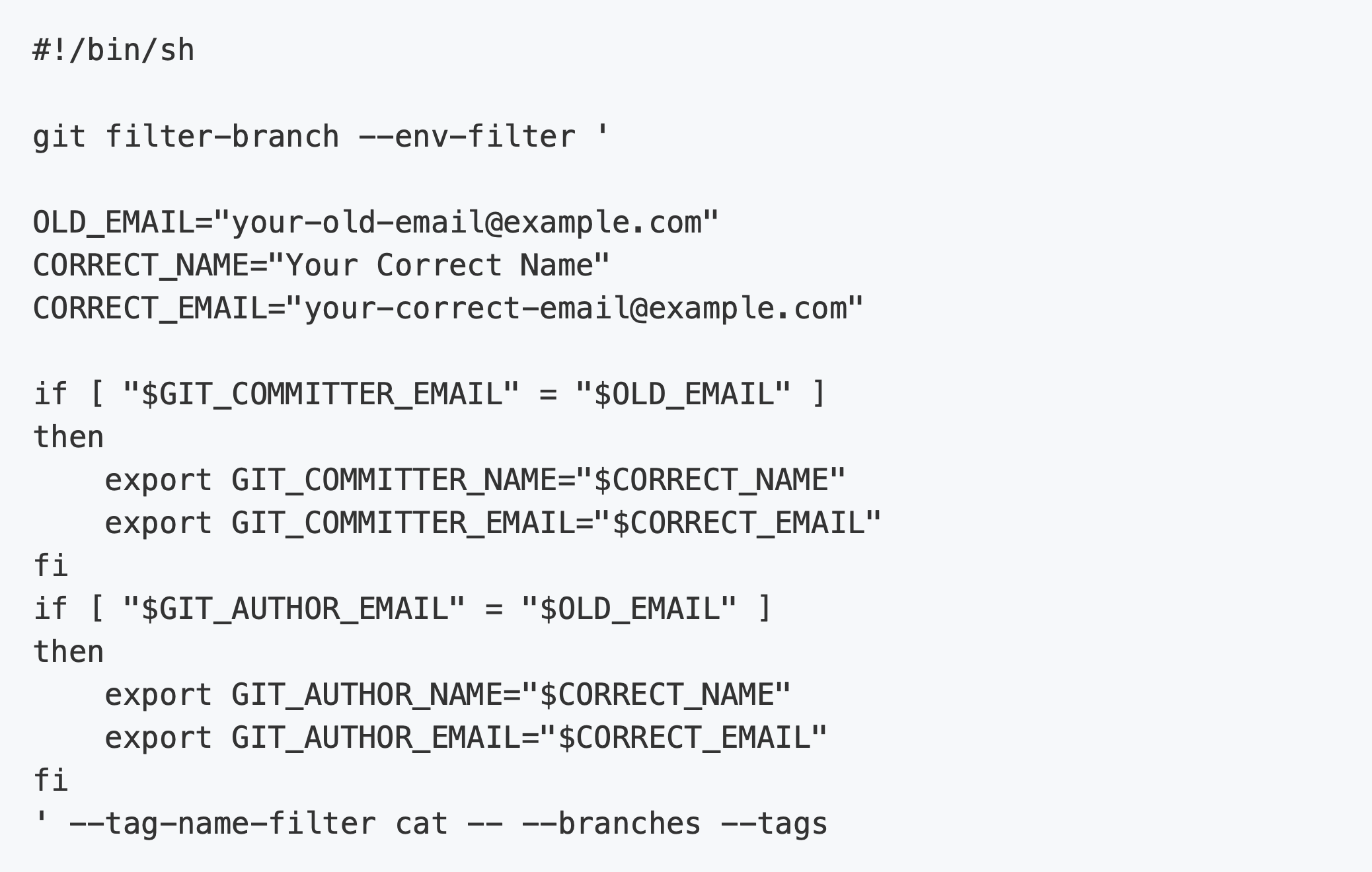}
	\caption[]{Script that changes the email address in every commit \footnotemark}
	\label{fig-fix-email-script}
\end{figure}

\begin{description}
	\item [Raise Awareness]

	An absolute must is to increase awareness among GitHub users about the threats that come by having their private email address publicly exposed along with valuable metadata that may be abused to create targeted phishing attacks. If users were transparently informed about this, the proportion of those that would still be tricked by a targeted social engineering attack would certainly decrease significantly. Also, users may voluntarily opt to use features like 2-Factor-Authentication or GitHub's \textit{noreply}-email-address service.
		
	\item [Require 2-Factor-Authentication]

	As described earlier, 2-Factor-Authentication probably isn't popular enough to protect a relevant number of accounts. Therefore GitHub should introduce strategies to voluntarily get people to use this feature -- one, just mentioned above, is to to simply raise awareness about it, clarifying about the potential damage when private, sensitive software leaks.
	
	More strictly, GitHub should also introduce a requirement for 2FA under certain circumstances. Metrics that can be used to determine the need of an account to use 2FA could be:
	
	\begin{itemize}
		\item Frequent activity
		\item Push access to multiple / popular repositories
		\item Part of private organizations
		\item Many private repositories
		\item Push access to other people's private repositories
	\end{itemize}
	
	Another approach would be to blame users that don't use 2FA. People inviting fellow developers to contribute to their potentially private repository would then see a warning badge, informing them that the user they are about to give push access has no advanced account protections put in place.
	
	Such a blaming badge could also be publicly displayed on the profile. It could then be combined with the metrics discussed above, so that only users with significant privileges and great potential of abuse in case of credential compromise would be blamed for not protecting their account properly.
	
	Of course, publicly disclosing whether a user hasn't 2FA enabled also poses a certain threat, but when only displaying the blaming badge in certain contexts (e.g. when inviting a user to collaborate), this threat may be negligible.
	
	\item [Make Noreply-Email-Addresses A Default]
	
	As the numbers presented in section \ref{monster-measurements} show, few people actually use Github's \textit{noreply}-email-addresses. While enabling this feature is rather unlikely to really protect users that have been actively contributing using a personal email address for a while, it can be fully effective when enabled for new users.
	
	This is why we suggest to enable this feature by default for new users, even going as far as blocking command line pushes that expose private email addresses by default.
	
	\item [Change Noreply-Email-Addresses]
	
	When collecting \textit{noreply}-email-addresses, an attacker can always immediately extract the GitHub username of the person because it is contained inside the \textit{noreply}-email-address. As this GitHub username is valuable information for the attacker, it would be safer to update the structure of the \textit{noreply}-email-addresses, by either:
	\begin{itemize}
		\item Just containing the user's GitHub ID (e.g. \texttt{1024025\allowbreak @users.noreply.github.com})
		\item Containing a random alphanumeric string, which can be resolved to the username by a GitHub API call.
	\end{itemize}
	
	Both of these options make it harder for the attacker to extract the GitHub username from the \textit{noreply}-email-address because they have to make one API call per user. Assuming API rate limits to be in place, this makes an automated phishing attack harder, while not impossible.
		
	\item [Automatic Cleanup]

	GitHub should introduce multiple automatic cleanup features. Those features would work in different domains, but share the same basic functionality: recreate commits that expose private email addresses and replace these email addresses with \textit{noreply}-email-addresses.
	
	There's certainly a danger of going too far: Integrity is an important promise of Git and if there exist services that automatically disturb this integrity by dropping and recreating modified versions of existing commits, that's a severe intervention with the way Git is expected to work. However, users can already do such things manually, e.g. by using the \texttt{filter-branch} command (see figure \ref{fig-fix-email-script}) and force-pushing thereafter. So, if a user knows the consequences, giving them multiple convenient options to have these kinds of things done automatically is certainly legitimate.
	
	Here are some ideas for automatic cleanup features:
	
	\begin{itemize}
		\item Recreate a cleaned version of an entire repository
		\item Clean a repository in-place
		\item Automatically clean pushed commits exposing private email addresses
	\end{itemize}
	
	Of course, when a user opts to enable such features, they may have to discard their local commit \textit{"versions"} when pulling even if they just created and pushed them.
	
	\item [Improve Push Blocking]
	
	Currently, the push blocking feature of GitHub is limited to blocking commits that use any email address associated with an GitHub account that has the blocking feature enabled. However, as some people may prefer to create their own \textit{noreply}-email-address, e.g. \texttt{commits@myowndomain.com}, and subsequently associate these email addresses with their GitHub account, the push blocking feature should have more configuration options. People preferring to have the \textit{noreply}-email-address under their control could then still block accidental pushes with their private email address, while allowing pushes with their dedicated commit email address.
	
	\item [Clone Rate Limit]
	
	It's the essence of a Git remote service to always be available as a remote. This is why there are no limits to repository cloning currently. At least, we didn't experience any, even when batch-cloning repositories in a concurrent manner.
		
	However, introducing a clone rate limit would create a relevant barrier to bots batch-cloning huge amounts of repositories and therefore significantly mitigate their effect. In addition to that, implementing a smart abuse detection mechanism is definitely suitable to some extent. We could imagine a mechanism that prevents further cloning from a specific IP address until a captcha is manually solved online.
		
	Again, such approaches may of course be circumvented by structures like botnets. But unfortunately, as is the case with encryption, you cannot fully protect against malicious actors by simply locking them out entirely. Instead, the goal is to make it incredible hard for them do perform their hacks, e.g. by requiring a disproportionate amount of time that's not feasible anymore.
\end{description}

\footnotetext{\cite{changing-author-info}}

\subsubsection{User}

\begin{description}
	\item [Noreply-Email-Address]

	Every GitHub user should either use a dedicated commit email address or GitHub's \textit{noreply}-email-address service, also enabling the option to block accidental command line pushes. In no case should private email addresses be used for Git commits. In every instance where they use Git, the user should take care of configuring a proper email address.
	
	If possible, existing commits should be overwritten with commits that use a protected email address. Of course, that's only possible in a limited scope, e.g. when a repository only has few contributors and no relevant forks / open PRs.
	
	\item[2-Factor-Authentication]
	
	Every GitHub user should have 2-Factor-Authentication enabled, especially if the user's privileges go beyond managing their very own repositories. It's not responsible not to do one's best to protect one's account if other people's data was affected in the case of an account compromise.
	
	\item[Raise Awareness]
	
	Beyond personally noticing the threat of public email address exposure that continues to exist even when \textit{noreply}-email-addresses are used from now on, and taking the required actions, it's the duty of developers aware of this issue to inform their colleagues about it. Especially beginners with Git should be guided to protect their personal email address right from the beginning.
\end{description}

\pagebreak

\section{Conclusion}

In this paper, we described an attack that first abuses GitHub's API to get a multitude of urls of repositories which are likely to have many contributors, then clones them and analyzes them using an advanced merging algorithm (see chapter \ref{attack}). The resulting database contains valuable information, e.g. allowing a malicious actor to create targeted phishing attacks. In chapter \ref{impact}, we presented real measurements of an actual implementation of this attack, referred to as \texttt{The Monster}. We were able to prove the great impact and the great threat originating from such an attack as a whole.

The need for advanced preventive measures against such attacks is therefore obvious. This is why, in chapter \ref{countermeasures}, after first showing the ineffectivity of the existing countermeasures as presented in chapter \ref{introduction}, we then proceeded to propose more effective countermeasures. We consider these preventive measures proposals the key part and main conclusion of this paper. For the problem being this relevant, we expect the responsible actors to install these proposed countermeasures as soon as possible as a mean to defeat further such attacks as presented in this paper.

Apart from that, it's our intention to spread the word about the abuse potential that lays with Git's email-address-based identity management when used in conjunction with public Git hosting services like GitHub. We call upon everyone in the industry with knowledge about this issue to inform fellow developers about the risk of exposing private email addresses in Git commits published publicly.

\textbf{Update (August 14, 2019):} As of today, a month after reaching out to GitHub, there are no changes. Also, GitHub has declined to address anything we proposed apart from considering ways to get more users to use GitHub's \textit{noreply}-email-address service.

\clearpage
\fancyhead[L]{}

\pagenumbering{Roman}
\setcounter{page}{\theromanpage}

\printbibliography[title={References}]
\pagebreak

\end{document}